\documentclass[runningheads]{cl2emult}

\usepackage{makeidx}  
\usepackage{graphicx} 
\usepackage{subeqnar} 
\usepackage{multicol} 
\usepackage{cropmark} 
\usepackage{lnp}      
\makeindex            

\usepackage{latexsym}

\begin{document}
\title*{An Experimental Study of a Granular Gas\protect\newline Fluidized by Vibrations}
\toctitle{An Experimental Study of a Granular Gas Fluidized\protect\newline by Vibrations}
%
%
\titlerunning{An Experimental Study of a Granular Gas Fluidized by Vibrations}
%
\author{{\'E}ric Falcon\inst{1}
\and St{\'e}phan Fauve\inst{1}
\and Claude Laroche\inst{2}}
\authorrunning{{\'E}ric Falcon et al.}
%
%
\institute{Laboratoire de Physique Statistique, {\'E}cole Normale
Sup{\'e}rieure, 24, rue Lhomond, 75231 Paris Cedex 05, France
\and Laboratoire de Physique, {\'E}cole Normale Sup{\'e}rieure de Lyon, 46
all{\'e}e d'Italie, 69364 Lyon Cedex 07, France}

\maketitle              

\begin{abstract}
\index{abstract}We report experimental results on the behavior of an ensemble of inelastically colliding particles, excited by a vibrated piston in a vertical cylinder. When the particle number is increased, we observe a transition from a regime where the particles have erratic motions (granular \lq\lq gas\rq\rq) to a collective behavior where all the particles bounce like a nearly solid body. In the gaslike regime, we measure the pressure at constant volume, and the bed expansion at constant external pressure, as a function of the number $N$ of particles. We also measure the density of particles as a function of the altitude, and find that the \lq\lq atmosphere\rq\rq\ is exponential far enough from the piston. From these three independent measurements, we determine a \lq\lq state equation\rq\rq\ between pressure, volume, particle number and the vibration amplitude and frequency.
\end{abstract}

\section{Introduction}
Once fluidized a vibrated granular medium looks like a gas of particles that can be described using kinetic theory of usual gases. However, granular gases basically differ from ordinary ones mostly due to the inelasticity of collisions, i.e., nonconservation of energy. While over the years many attempts based on kinetic theory \cite{Jenkins83} have been made to describe such dissipative granular gases, no agreement has been found so far both with experiments \cite{Warr95,Luding94a} and numerical simulations \cite{Luding94a,Luding94b,Luding95,Herrmann98}, for the dependence of the \lq\lq granular temperature\rq\rq, i.e., the mean kinetic energy per particle, on the parameters of vibration \cite{Lee95,McNamara98,Kumaran98,Huntley98}. The aim of this study is to guess possible gas-like state equations for such a dissipative granular gas and to observe new kinetic behaviors which trace back to the inelasticity of collisions, e.g., the tendency of such media to form clusters. Although this feature has probably been known since the early observation of planetary rings \cite{Goldreich82} and although various cluster types in granular flows have been observed numerically \cite{Hopkins91}, there exist only a few recent laboratory experiments. One experiment, with a horizontally shaken 2--D layer of particles, displayed a cluster formation, but the coherent friction force acting on all the particles was far from being negligible \cite{Kudrolli97}. We performed a similar experiment by exciting a 3--D granular medium with a vertically vibrated piston. We did observe clustering, but we could not rule out a lock-in mechanism involving the time scale connected with gravity and the period of vibration \cite{Falcon99c}. We thus repeated this experiment in a low-gravity environment, where inelastic collisions were the only interaction mechanism, and observed a motionless dense cluster that confirms that the inelasticity of collisions alone can generate clustering \cite{Falcon99b}.

This paper is devoted to the study of the low-density situation, where clustering does not occur. It is organized as follows. Section 2 is devoted to the presentation of our experimental setup. The experimental results are presented in Secs. 3--5. We report in Sec. 3 (resp. Sec. 4) the measurements of pressure (resp. volume) of a gas of spherical particles excited by a vibrating piston and undergoing inelastic collisions. At constant external driving, we show that the pressure passes through a maximum for a critical number of particles before decreasing for large $N$. The density of particles as a function of the altitude is studied in Sec. 5, where we observe an exponential atmosphere far enough from the piston. From measurements of Secs. 3--5, we show in Sec. 6 that the dependence of the \lq\lq granular temperature\rq\rq, $T$, on the piston velocity, $V$, is of the form $T \propto V^{\theta}$, where $\theta$ is a decreasing function of $N$. Finally, we discuss our results in the light of previous works \cite{Warr95,Luding94a,Luding94b,Luding95,Herrmann98,Lee95,McNamara98,Kumaran98,Huntley98} and give our conclusions in Sec. 7.  

\section{Experimental Setup}
The experiment consists of a transparent cylindrical tube, with an inner diameter of $60$\ mm, filled from $20$ up to $2640$ stainless steel spheres, $2$\ mm in diameter, roughly corresponding to $0$ up to about $5$ particle layers at rest. An electrical motor, with eccentric transformer from rotational to translational motion, drives the particles sinusoidally with a $25$ or $40$ mm amplitude, $A$, in the frequency range from $6$ to $20$ Hz. A lid in the upper part of the cylinder, is either fixed at a given height, $h$ (constant--volume experiment) or is mobile and stabilized at a given height $h_{m}$ due to the bead collisions (constant--pressure experiment). The heights $h$ and $h_{m}$ are defined from the lower piston at full stroke.

\begin{figure}
\centering
\includegraphics[width=.71\textwidth]{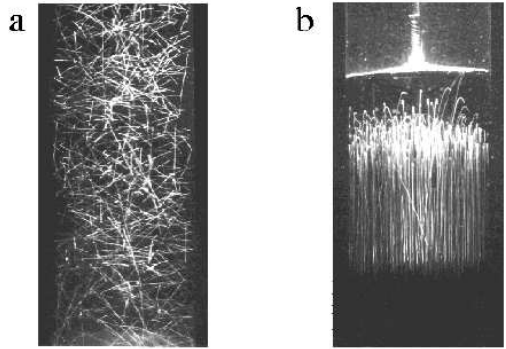}
\caption[]{Transition from a dissipative granular gas to a dense cluster: ({a}) $N= 480$; ({b}) $N=1920$, respectively corresponding to roughly $1$ and $4$ particle layers at rest. The parameters of vibration are $f = 20$\ Hz and $A = 40$\ mm. The driving piston is at the bottom (not visible), the inner diameter of the tube being $52$ mm}
\label{fig01}
\end{figure}

When the vibration is strong enough and the number of particles is low enough, the particles display ballistic motion between successive collisions like molecules in a gas (see Fig.\ \ref{fig01}a). When the density of the medium is increased, the gaslike state is no longer stable but displays the formation of a dense cluster bouncing like a nearly solid body (see Fig.\ \ref{fig01}b). This paper is devoted to the study of the gaslike state. Its aim is to determine experimentally a \lq\lq state equation\rq\rq\ between pressure, volume, particle number and the vibration parameters (amplitude and frequency).

\section{Pressure Measurements}
\begin{figure}
\centering
\includegraphics[width=.6\textwidth]{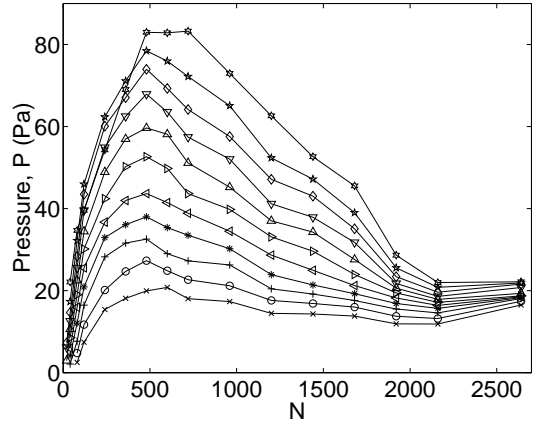}
\caption[]{Mean pressure $P$ as a function of $N$. From the lower ($\times $--marks) to the upper (hexagrams) curve, vibration frequency $f$ varies from $10$ to $20$\ Hz with a $1$\ Hz step. For all these experiments, $h-h_{0}= 5$ mm and $A = 25$\ mm, $h_{0}$ being the bed height at rest. One single layer of particles at rest corresponds to $N=600$. Lines join the data points}
\label{fig02}
\end{figure}

Time averaged pressure measurements have been performed as follows. Initially, a counterwheight of mass $46$\ g balances the lid mass. The piston drives stainless steel spheres in erratic motions in all directions (see Fig.\ \ref{fig01}a). Particles are hitting the lid all the time, so that to keep it at a given height $h$ we have to hold the lid down by a given force, $Mg$, where $M$ is the mass of a weight we place on the lid and $g$ the acceleration of gravity. At a fixed $h$, {\it i.e.} at a constant--volume, Fig.\ \ref{fig02} shows the time averaged pressure $P=Mg/S$ exerted on the lid as a function of the number $N$ of beads in the container, for different frequencies of vibration, $S$ being the area of the tube cross-section. At constant external driving, {\it i.e.} at fixed $f$ and $A$, the pressure passes through a maximum for a critical value of $N$ roughly corresponding to $0.8$ particle layers at rest. This critical number is independent of the vibration frequency. A further increase of the number of particles leads to a decrease in the mean pressure since more and more energy is dissipated by inelastic collisions. Note that gravity has a small effect in these measurements that are performed for $V^{2} \gg gh$, where $V=2\pi fA$ is the maximum velocity of the piston. For $N$ such that one has less than one particle layer at rest, most particles perform vertical ballistic motion between the piston and the lid. Thus, the mean pressure increases roughly proportionally to $N$. When $N$ is increased such that one has more than one particle layer at rest, interparticle collisions become more frequent. The energy dissipation is increased and thus the pressure decreases.

\begin{figure}
\centering
\includegraphics[width=.6\textwidth]{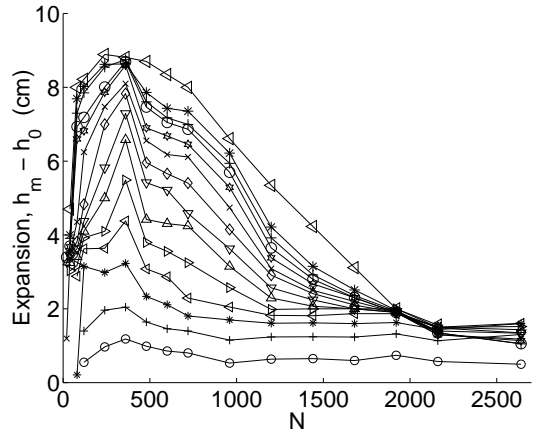}
\caption[]{Maximal bed expansion, $h_{m}- h_{0}$, as a function of $N$, for various frequencies $f$ of vibration. From the lower ($\circ$--marks) to the upper ($\lhd$--marks) curve, $f$ varies from $7$ to $20$\ Hz with a $1$\ Hz step and $A = 25$\ mm. One single layer of particles at rest corresponds to $N=600$. Lines join the data points}
\label{fig03}
\end{figure}

\section{Volume Measurements}

We now consider the bed expansion under the influence of collisions on a circular wire mesh lid placed on top of the beads leaving a clearance of about 0.5 mm between the edge of the lid and the tube. Due to the bead collisions, the lid is stabilized at a given height $h_{m}$ from the piston at full stroke. Although the lid mass is roughly $50$ times smaller than the total mass of the beads, the lid proves to be quite stable and remains horizontal. The expansion, $h_{m} - h_{0}$, of the bed is displayed in Fig.\ \ref{fig03} as a function of $N$ for different vibration frequencies. $h_{0}$ is the bed height at rest. At fixed $f$, the expansion passes through a maximum for a critical value of $N$ roughly corresponding to $0.6$ particle layers at rest. This critical number is independent of the vibration frequency. When $N$ is further increased, the expansion decreases showing, as for pressure measurements, an increase in dissipated energy by inelastic collisions. Note that the height $h_{m}$ of the granular gas is much larger than for the pressure measurements of Fig.\ \ref{fig02}. Consequently, gravity is important here.

\begin{figure}
\centering
\begin{tabular}{cc}
\includegraphics[width=.49\textwidth]{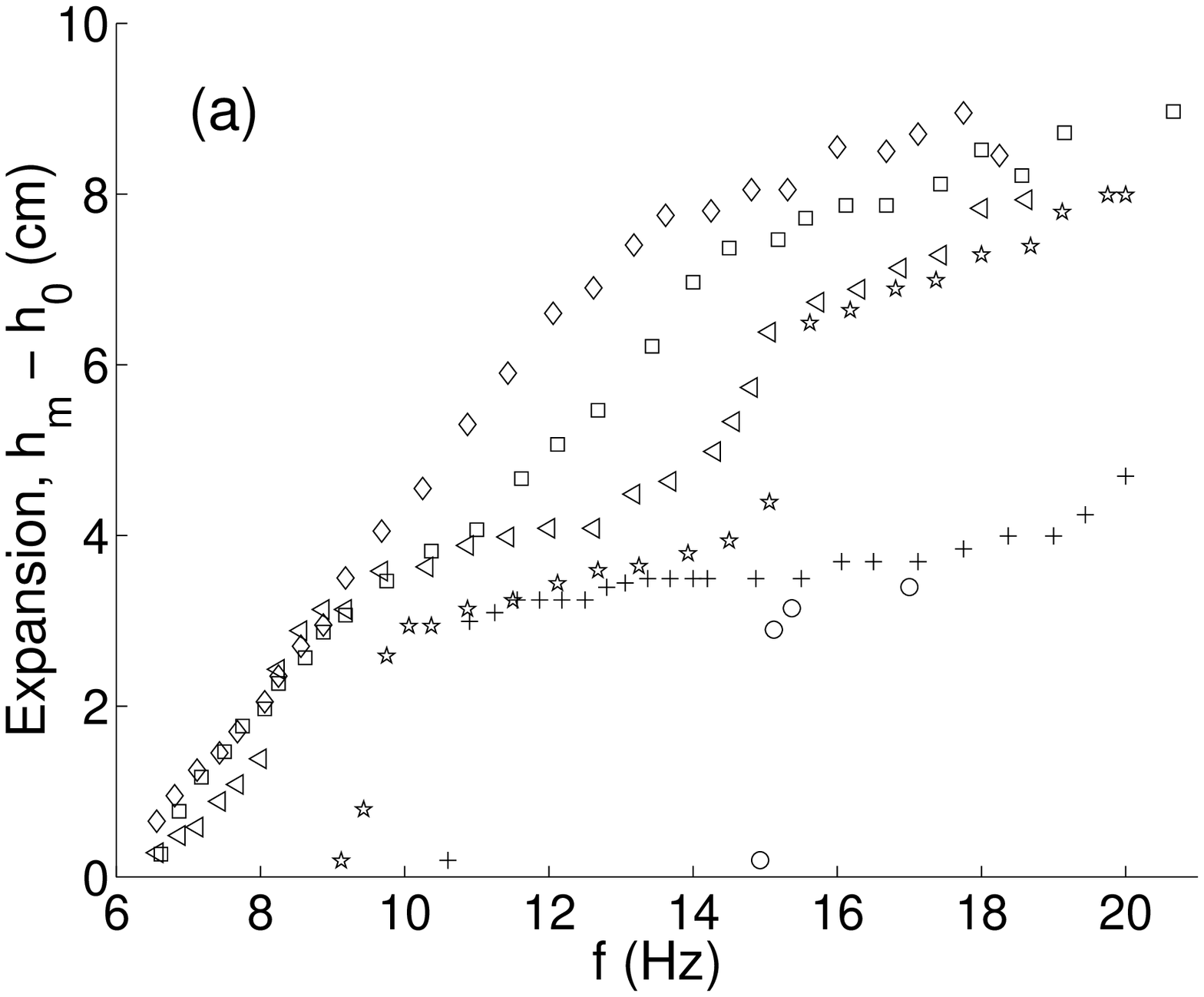}
&
\includegraphics[width=.49\textwidth]{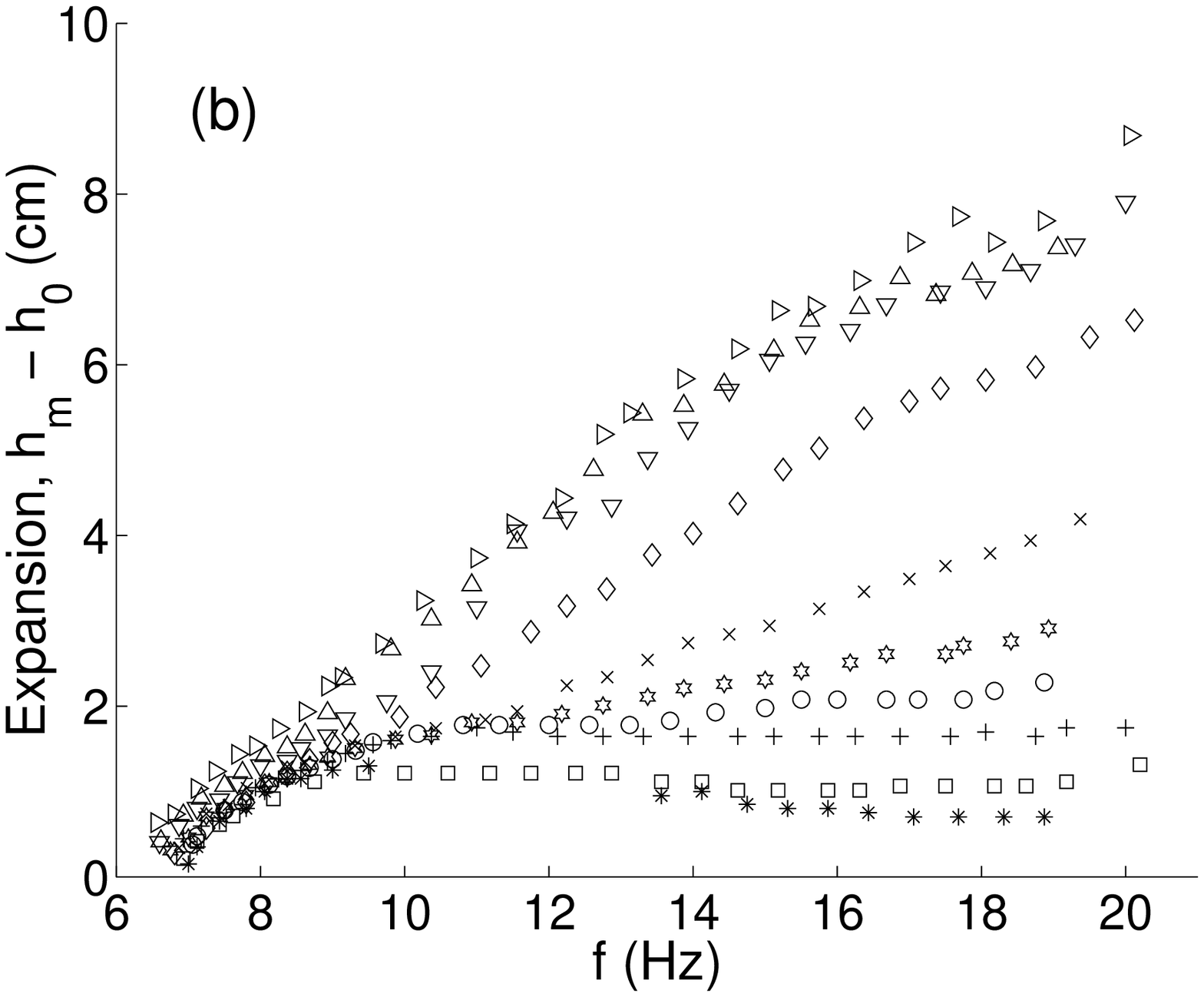}
\end{tabular}
\centering
\caption[]{Maximal bed expansion, $h_{m}- h_{0}$, as a function of $f$, for various numbers of particles $N$. ({a}) From the {\em lower} ($\circ$--marks) to the {\em upper} ($\Diamond$--marks) curve, $N= 20$, $40$, $80$, $120$, $240$ and $360$. ({b}) From the {\em upper} ($\rhd$--marks) to the {\em lower} ($\ast$--marks) curve, $N= 480$, $600$, $720$, $960$, $1200$, $1440$, $1680$, $1920$, $2160$ and $2640$. For all these experiments $A = 25$ mm. One single layer of particles at rest corresponds to $N=600$}
\label{fig04}
\end{figure}

As already found experimentally in 1--D \cite{Luding94a} and 3--D \cite{Brennen96,Hunt94} and numerically \cite{Luding94a,Lan95}, we observe that, at $N$ fixed, the granular medium exhibits (see Fig.\ \ref{fig04}a) a sudden expansion at a critical frequency corresponding to a bifurcation similar to that exhibited by a single ball bouncing on a vibrating plate \cite{Luding94a,Brennen96}. Moreover, Fig.\ \ref{fig04}a shows that this critical frequency depends on the number of layers, $n$. When $n$ increases above 0.4, a transition from the 1--D--like behavior to a 3--D one is observed: the expansion at the critical frequency becomes less abrupt and tends to increase regularly with $f$ (see Figs.\ \ref{fig04}a--b).

\section{Density Measurements}

\begin{figure}
\begin{center}
\includegraphics[width=.48\textwidth]{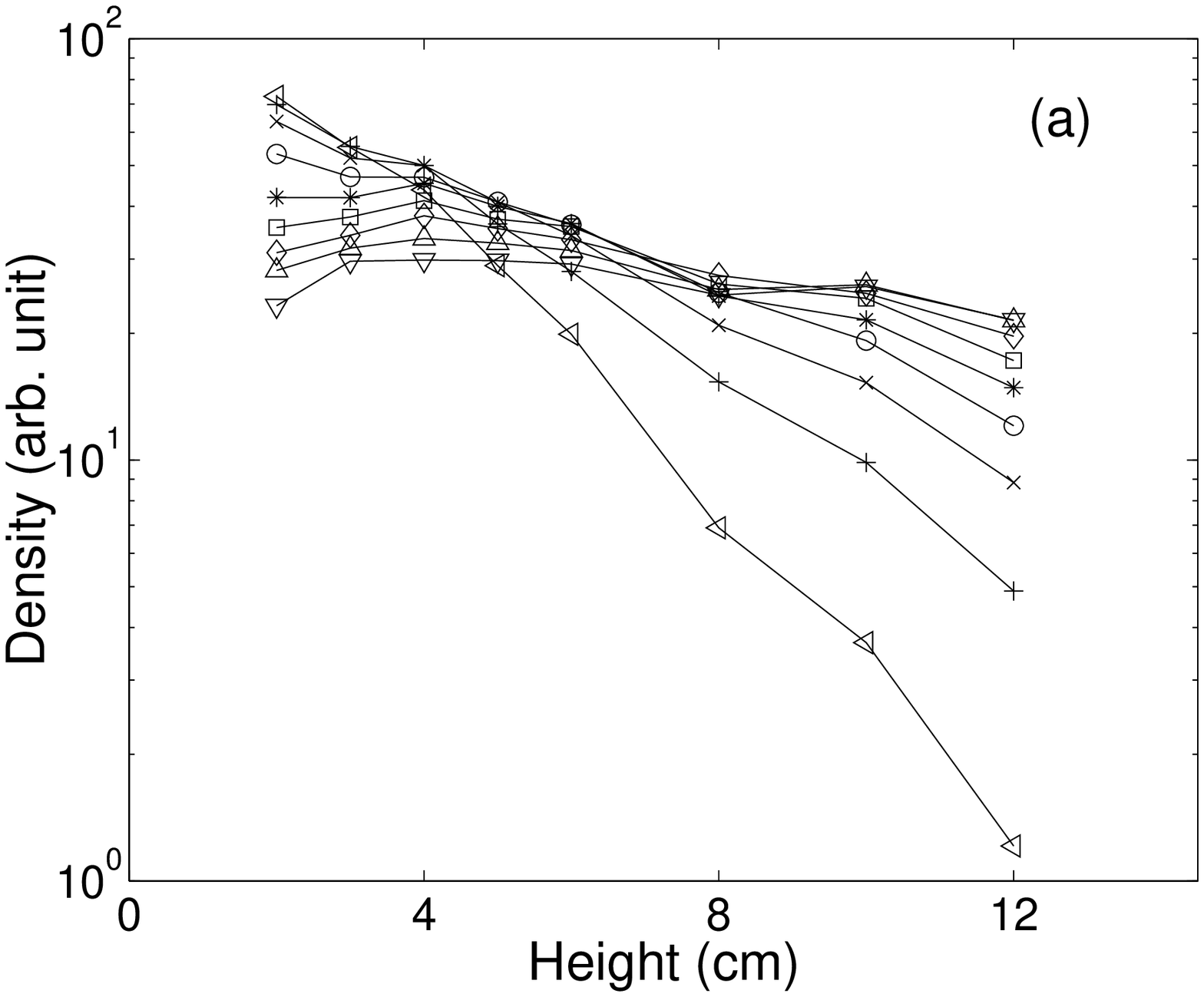}
\includegraphics[width=.48\textwidth]{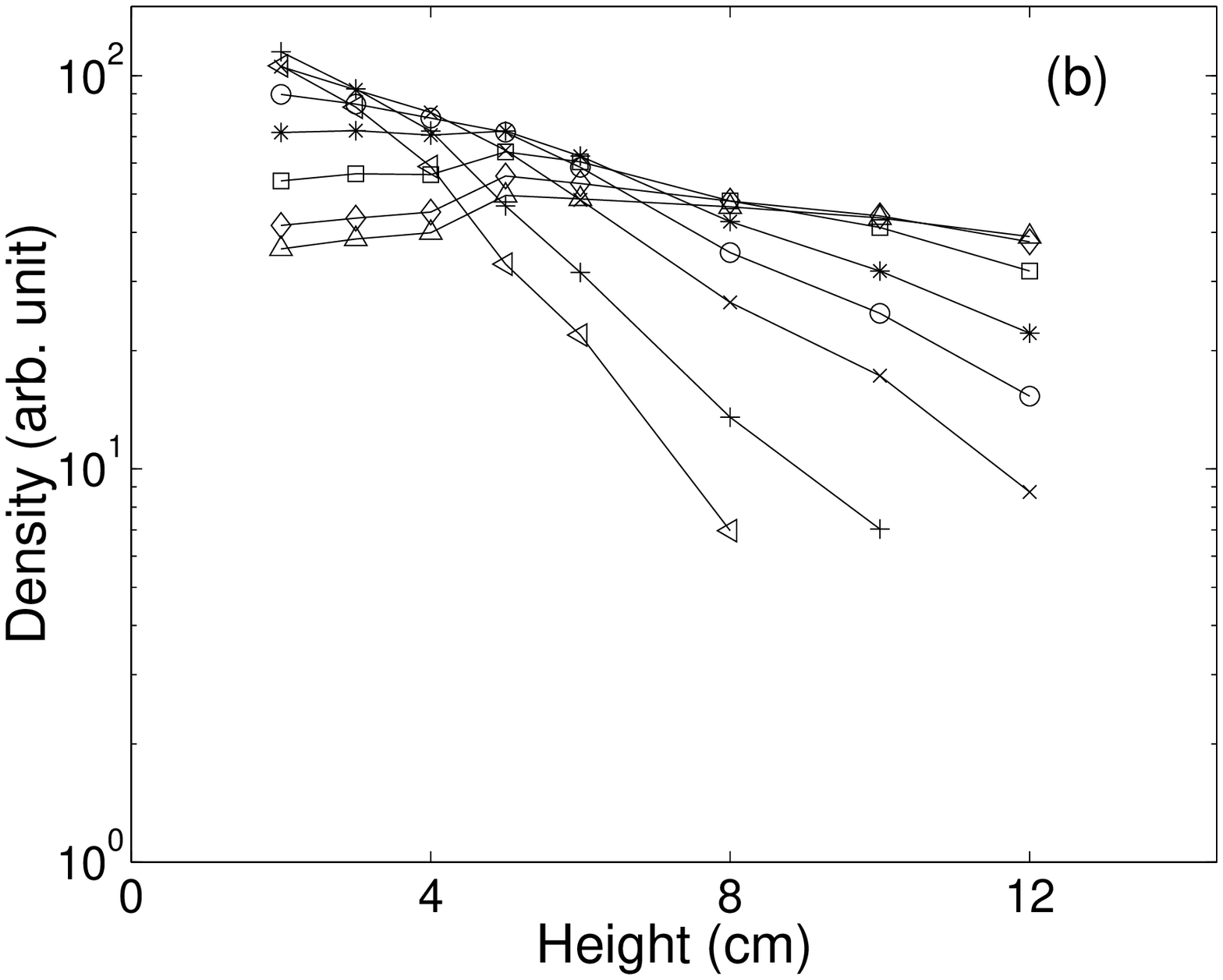}
\includegraphics[width=.48\textwidth]{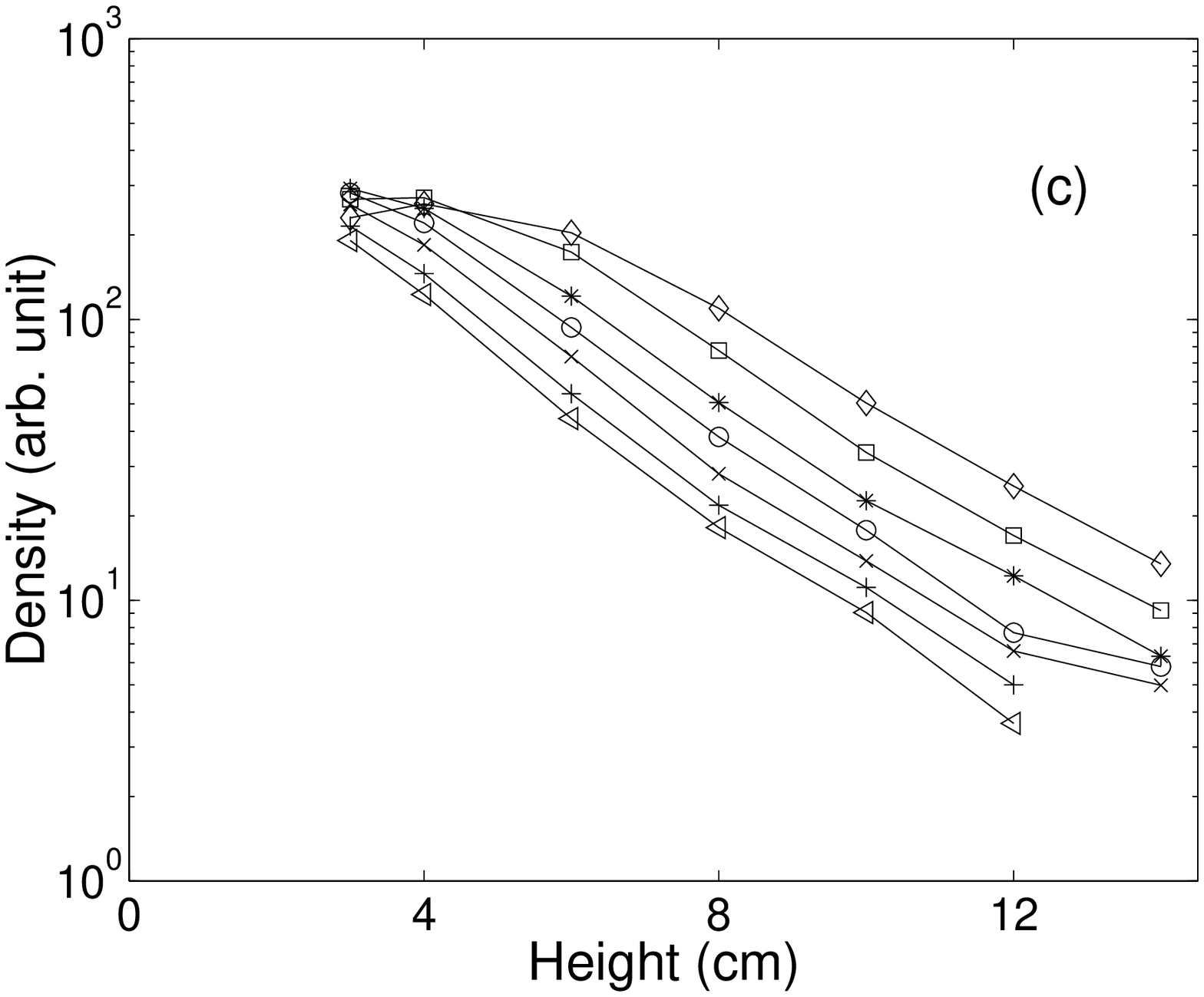}
\end{center}
\caption[]{Mean density as a function of the height, for various
frequencies $f$ of vibration and $3$ numbers of particles ({a}) $N=480$: $f=$ ($\lhd$) $5$, ($+$) $6$, ($\times $) $7$, ($\circ$) $8$, ($\ast$) $9$, ($\Box$) $10$, ($\Diamond$) $11$, ($\triangle$) $12$ and ($\bigtriangledown$) $13$\ Hz; ({b}) $N=720$: $f=$ ($\lhd$) $5$, ($+$) $5.6$, ($\times $) $7.1$, ($\circ$) $8.6$, ($\ast$) $10.1$, ($\Box$) $12.2$, ($\Diamond$) $14$ and ($\triangle$) $15.2$\ Hz; ({c}) $N=1440$: $f=$ ($\lhd$) $7$, ($+$) $8$, ($\times $) $10$, ($\circ$) $11.4$, ($\ast$) $12.8$, ($\Box$) $15$ and ($\Diamond$) $17$\ Hz. For all these experiments, $A=40$\ mm. One single layer of particles at rest corresponds to $N=480$. Lines join the data points}
\label{fig05}
\end{figure}

Time averaged density measurements at a given height $z$ are performed by means of two closely coupled coils, $\delta z=5$ mm in height, and $64$\ mm in inner diameter, the cylindrical tube now being $52$ (resp. $62$)\ mm in inner (resp. outer) diameter. An 1.5 kHz a.c. voltage is applied to the primary coil, the turns ratio of the transformer being roughly equal to 2. Steel spheres moving accross the permanent magnetic field of the primary coil, generate an inductive voltage variation accross the secondary one. This root mean square a.c. voltage $\Delta U$ is a function of the mean inductance and mutual variations which are proportional to the mean number of particles in the volume delimited by the sensor at altitude $z$ from the piston surface at full stroke. We have calibrated the sensor with steel spheres at rest and we have checked that $\Delta U \propto N$. We have also found that the effect of spheres outside of the sensor volume decays exponentially with the distance to the sensor, with a $10 \pm 1$ mm decay length independent of $N$, for our range of $N$.  Particles density as a function of altitude is shown in Figs.\ \ref{fig05}a--c for 3 different total numbers of particles and for various frequencies. For each $N$, the profile density in log--linear axes displays a decay (at low $f$), a plateau (at intermediate $f$) or a dip (at high $f$) near the piston and an exponential decay in the tail at high altitude, whatever $f$.

\begin{figure}
\centering
\includegraphics[width=.56\textwidth]{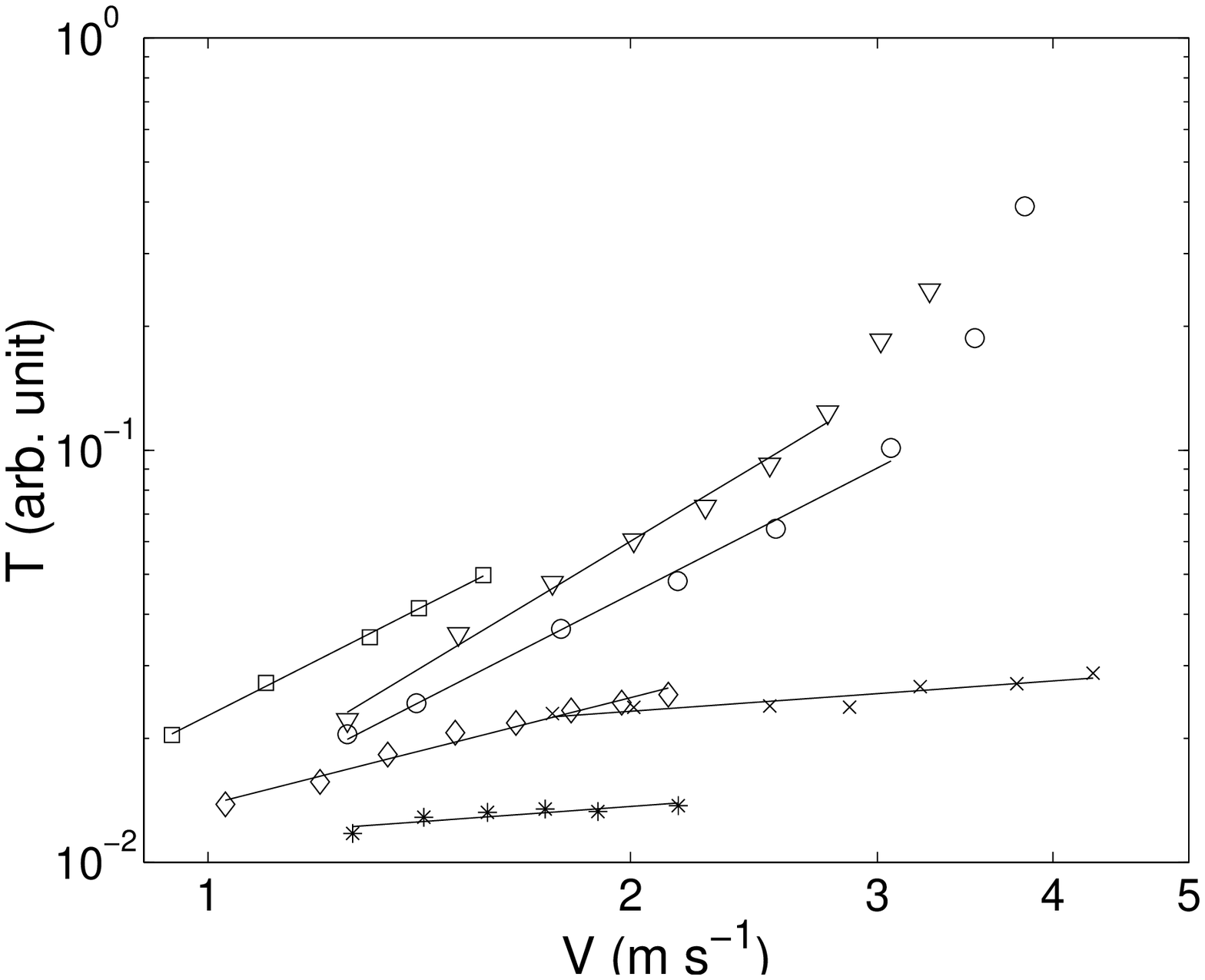}
\caption[]{Log-log plot of granular temperature versus $V$ for various numbers of layers: ($\bigtriangledown$) $0.8$, ({\small $\Box$}) $1$, ($\circ$) $1.2$, ({\small $\Diamond$}) $2$, ($\times $) $2.4$ and ($\ast$) $3$. Experiments ({\small $\Box$}), ({\small $\Diamond$}) and ($\ast$) (resp. ($\bigtriangledown$), ($\circ$) and ($\times $)) are performed for $A=25$ mm (resp. $A=40$ mm). Power law fits of the form $V^{\theta}$ are dispayed in solid lines}
\label{fig06}
\end{figure}

As for an isothermal gas, the atmosphere is found to be exponential far enough from the piston, but on very different length scales, i.e., few cm (resp. km) for our experiment (resp. for air). Such a dense upper region supported on a fluidized low-density region near the piston has been also reported numerically \cite{Lan95} and predicted theoretically \cite{Kurtze98}. Although the dip in the density profiles at the bottom was already observed in a 2--D granular gas experiment, non-negligible coherent friction force acting on all the particles did not allow determination of the granular temperature dependence on the piston velocity $V$ from exponentional Boltzmann distributions fitted to tails of density profiles \cite{Warr95}. We can fit an exponential curve to the tail of the profile density. From the decay rate, $\xi$, in the fitted exponential and using kinetic theory \cite{Warr95}, we can extract the dependence of the granular temperature $T$ on the piston velocity. In Fig.\ \ref{fig06}, we plot $-1/\xi$ (which is proportional to $T$) against $V$ with log-log-axis together with power law fits of the form $T \propto V^{\theta}$ where $\theta$ is $n$--dependent. Note that this power law, being observed only on a small range of velocities, one cannot rule out another functional behavior. In particular, the faster increase of $T$ at high velocity is not significant because of imprecision on the exponential decay of the density (see Fig.\ \ref{fig05}).

\begin{figure}
\centering
\includegraphics[width=.56\textwidth]{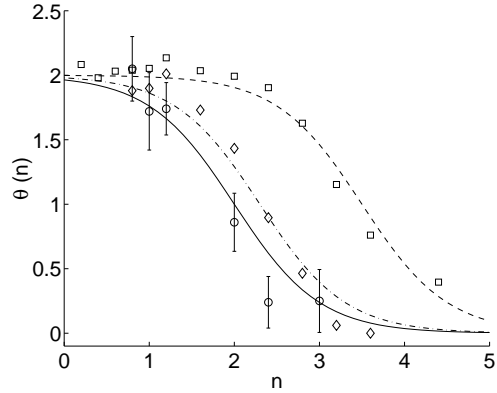}
\caption[]{Evolution of the exponent $\theta$ as a function of the number of layers, $n$, from pressure ({\small $\Box$}), volume ({\small $\Diamond$}) and density ($\circ$) measurements. Fits are $\theta(n) = 1 - \tanh (n-n_{c})$ with $n_{c} = 2$ ({\it straight line}), $2.3$ ({\it dot-dashed line}) and $3.5$ ({\it dashed line})}
\label{fig07}
\end{figure}

\section{Towards a State Equation}

In order to use the above measurements to determine a state equation, we have to find the appropriate dependence $T = T(V, n)$ of the granular temperature as a function of the vibrating velocity $V$ and the number of particle layers $n$. It is known that for a fixed number $n$ of granular layers at rest, one has in the low-density limit $P\Omega \propto T$ \cite{McNamaraPrivate}, where $P$ is the mean pressure and $\Omega$ the volume. Taking into account this law for small densities and our observation $T \propto V^{\theta(n)}$ from density measurements in an exponential atmosphere, we have plotted $\log (P)$ and $\log (h_{m} - h_{0})$ as functions of $\log (V)$. On the reported frequency range, these curves are straight lines, the slopes of which give $\theta(n)$. The behavior of $\theta(n)$  for the experiments at constant volume in Sec. 3 (resp. constant pressure in Sec. 4) is displayed in Fig.\ \ref{fig07} with {\small $\Box$}--mark (resp. {\small $\Diamond$}--mark) together with the one in $\circ$--mark extracted from exponential density profiles of Sec. 5. The three curves, obtained with different experimental conditions and independent measurements have the same shape which could be simply fitted by $\theta = 1 - \tanh (n - n_{c})$ where $n_{c} = 3.5$ (resp. 2.3) and 2.

\begin{figure}
\centering
\begin{tabular}{cc}
\includegraphics[width=.49\textwidth]{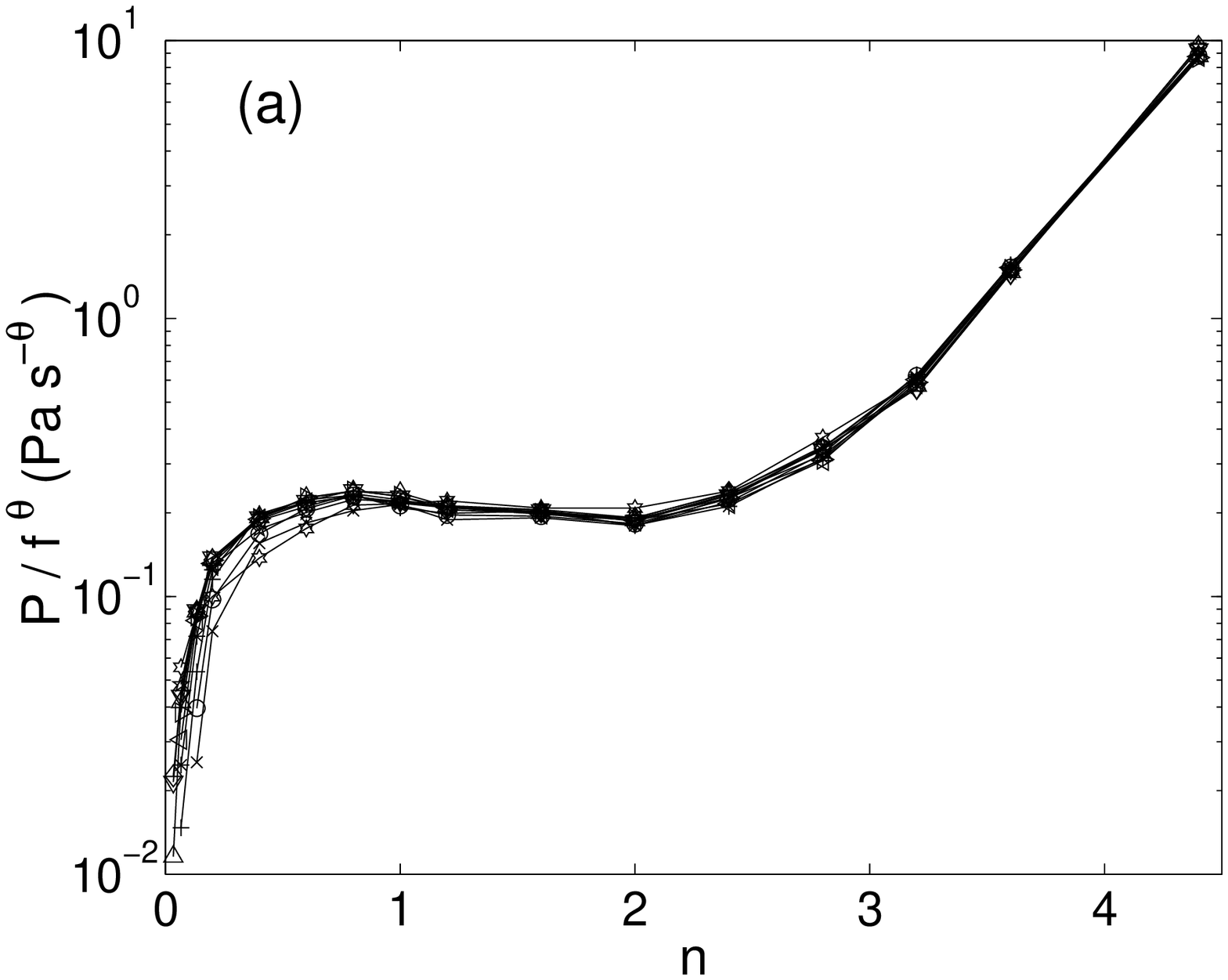}
&
\includegraphics[width=.49\textwidth]{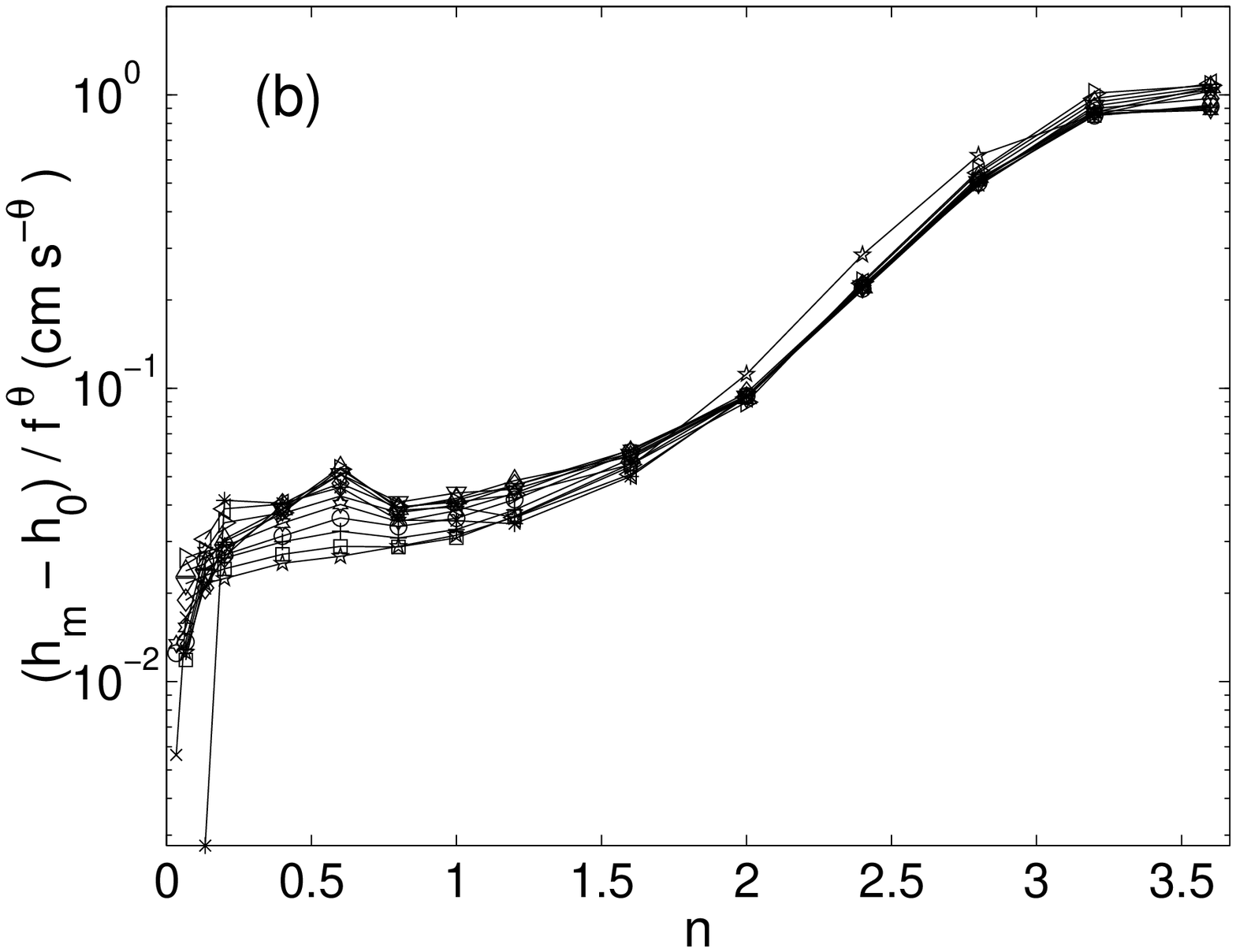}
\end{tabular}
\centering
\caption[]{({a}) Mean pressure $P$ from Fig.\ \ref{fig02} rescaled by $f^{\theta}$ as a function of $n$. $\theta(n) = 1 - \tanh (n-n_{c})$ with $n_{c} = 3.5$. ({b}) Maximal bed expansion, $h_{m} - h_{0}$, from Fig.\ \ref{fig03} rescaled by $f^{\theta}$ as a function of $n$. $\theta(n) = 1 - \tanh (n-n_{c})$ with $n_{c} = 2.3$}
\label{fig08}
\end{figure}

We can now use the observed law $T \propto V^{\theta(n)}$ to scale the pressure and bed expansion measurements of Figs.\ \ref{fig02} and \ref{fig03}. The results are displayed in Figs.~\ref{fig08}a--b and show a rather good collapse of all the data on a single curve. We have thus shown that the law, $P\Omega \propto T$, together with $T \propto V^{\theta(n)}$, provide a correct empirical state equation for our dissipative granular gas in the kinetic regime. As shown earlier, this regime is limited at high density by the clustering instability \cite{Kudrolli97,Falcon99c,Falcon99b} and on the other side, for a fixed too small number of particles, when the gas suddenly contracts on the piston below a critical frequency (see Fig.\ \ref{fig04}a and \cite{Luding94a,Brennen96,Lan95}).

\section{Conclusion}
The aim of this study was an experimental determination of the state equation of dissipative granular gases. It is known that for a fixed number of granular layers at rest, one has in the low-density limit $P\Omega \propto T$ \cite{McNamaraPrivate}, where $P$ is the mean pressure, $\Omega$ the volume and $T$ the \lq\lq granular temperature\rq\rq. However, the dependence of $T$ on the vibration amplitude, $A$, and frequency, $f$, of the piston and on the number of particles, $n$, is still a matter of debate \cite{Lee95,McNamara98,Kumaran98,Huntley98}. Kinetic theory \cite{Warr95,Kumaran98} or hydrodynamic models \cite{Lee95} show $T \propto V^{2} n^{-1}$, whereas numerical simulations \cite{Luding94a,Luding94b,Luding95,Herrmann98} or experiments \cite{Warr95,Luding94a} give $T \propto V^{\alpha} n^{-\beta}$, with $1.3 \leq \alpha \leq 2$ and $0.3 \leq \beta \leq 1$, where $V=2\pi fA$ is the maximum velocity of the piston. These previous conflicting results may be explained by the $V$--dependence of $T$ that we have found from our measurements: $T \propto V^{\theta(n)}$, with $\theta$ continuously varying from $\theta = 2$ when $n \rightarrow 0$, as expected from kinetic theory, to $\theta \simeq 0$ for large $n$. We emphasize that we have only considered the dependence of the granular temperature on the vibration velocity $V$. For the dependence on the number of particles, another term of the form $n^{-\beta}$ should exist, that we cannot determine from these measurements.


\end{document}